\documentstyle[amstex,floats,twocolumn,prl,aps]{revtex}

\begin{document}


{\bf Comment on ''Minimal Surfaces, Screw Dislocations, and Twist Grain
Boundaries''}

\bigskip 

In a recent Letter,  Kamien and Lubensky \cite{Lub} considered the
arrangement of screw dislocations in the Twist Grain Boundaries (TGB) in
layered structures. In particular, they calculated the energy of the
surface constructed via a linear superposition of screw dislocations (LSD)
in Smectic-A (SmA) phase, parallel to ${\bf z}$ with separation $l_{d}$
along ${\bf y}$ axis. The authors concluded that the leading contribution to
the interaction energy $E_{int}$ between $k-$quanta dislocations is cutoff
dependent through the logarithmic term $\sim b^{4}l_{d}^{-2}\ln (l_{d}/\xi )$
where the cutoff \ ''size of the dislocation core'' $\xi $ is of the order
of the inter-layer spacing $d$ and $b=kd$ is the Burger's vector. The
positive sign of interaction coresponds to {\it repulsion} between
dislocations and therefore, according to \cite{Lub}, the system with larger
dislocation separation will have a lower energy.

In the present Comment we would like to object that the long-range interaction
between dislocations in array is {\it attractive}. This  attraction
combined with the short-range repulsion we justify below, results in the
nonmonotonous behavior of $E_{int}(l_d)$ and can provide the (local)
stability of the dislocation array at finite $l_d$. This conclusion can be
important to explain the discontinuity of the
TGB - SmA phase transition in chiral liquid crystals where due to specific
chiral intermolecular interaction the dislocation energy is negative
and dislocation arrays penetrate inside the sample until the interaction
between them stabilizes the energy win \cite{TGBtheor}.

To find the interaction explicitly we repeat the calculations of \cite
{Lub} presenting the elastic energy of the distorted SmA in
the more convenient complex variables $(\zeta ,\overline{\zeta })=x\pm iy$
and $(\partial ,\overline{\partial })=(\partial _{x}\mp i\partial _{y})/2$
as: 
\begin{equation}
F=K(\partial \overline{\partial }u)^{2}+\frac{B}{2}[\partial
_{z}u-2(\partial u)(\overline{\partial }u)]^{2},  \label{Elast}
\end{equation}
where $u$ is the layers distortion. The energy of the single screw
dislocation with the distortion field $u_{d}=(b/2\pi )\arctan(y/x)$ $%
=i(b/4\pi )\ln (\overline{\zeta } /\zeta )$ and the stress $\partial
u_{d}=-ib/4\pi \zeta$; $\partial _{z}u=0$ can be easily calculated from (\ref
{Elast}). The bending (linear) part of the energy $(\overline{\partial }
\partial u)^{2}$ is equal to zero and the compression (nonlinear) part
diverges after integration as $2\pi B(b/4\pi )^{4}\xi ^{-2}$ \cite{Kleman}.

The stress $\partial u$ of the LSD array is formed from the individual
contributions of single dislocations as: 
\begin{equation}
\partial u_{LSD}=-i\frac{b}{4\pi }\sum_{n}\frac{1}{\zeta +inl_{d}}
=-i\frac{b}{4l_{d}}\coth \frac{\pi \zeta }{l_{d}}  \label{Wall}
\end{equation}

As pointed in \cite{Lub}$\,$,$\ $the uniform dilatation of SmA layers $%
2(b/4l_{d})^{2}z$ should also be performed. Otherwise the compression part
of (\ref{Elast}) diverges at $x\rightarrow \infty $. The LSD distortion
field  after integration of (\ref{Wall}) is written as: 
\begin{equation}
u_{LSD}=\frac{b}{4\pi }\left[ i\ln \frac{\sinh (\pi \overline{\zeta}/l_d) }
{\sinh (\pi \zeta /l_d)} +2(\frac{b}{4l_{d}})^{2}z\right] \end{equation}

Consider the elastic energy of the LSD array on the fundamental (periodic)
stripe region $\left| y\right| <l_{d}/2$ that contains one dislocation at $%
\zeta =0$. Similar to the single dislocation case, the bending contribution
vanishes and the compression part diverges at the dislocation core. To catch
this divergency we substitute $\partial u_{LSD}$, $\partial_z u_{LSD}$ into
(\ref{Elast}) and pick the singular terms of $F_{LSD}$ at $\zeta =0$:  
\begin{eqnarray}
&F&_{LSD}=2B(\frac{b}{4l_{d}})^{4}[1-\coth \frac{\pi \zeta }{l_{d}}
\coth \frac{\pi \overline{\zeta }}{l_{d}}]^{2}  \label{FLSD} \\
=2&B&(\frac{b}{4\pi })^{4}\frac{1}{(\zeta \overline{\zeta})^{2}}
-B\frac{b^{4}}{4(4\pi l_{d})^{2}}\frac{1}{\zeta \overline{\zeta }}
+2B(\frac{b}{4l_{d}})^{4}F_{R}(\zeta /l_{d}),  \nonumber
\end{eqnarray}
where $F_{R}$ is the regular part of the energy. Being integrated, the first
terms diverges as $\xi ^{-2}$ and corresponds to the self-energy of the
separate dislocation. The second term diverges as $\ln \xi $ and the third
one gives a constant. The two last terms depend on $l_{d}$ and can be
attributed to the interaction between dislocations. After integration, the
resulting interaction energy per one dislocation 
$E_{int}=Bb^{4}/(32\pi l_{d}^{2})[-\ln (l_{d}/\xi )+C]$  
 corresponds to the Kamien and Lubensky logarithmic interaction, but with the
{\it negative} sign. Therefore, dislocations attract each other when 
$l_{d}\gg \xi $ and repulse when $l_{d}\sim $ several $\xi $ \ when
$E_{int}$ changes the sign. Although the last statement lies beyond the
applicability of the LSD approximation $ l_{d}/\xi \gg 1$, one can justify
the short-range attraction by  that the two separated one-quanta dislocations
with energy $2E_{1}\sim 2\cdot 2\pi B(d/4\pi )^{4}\xi ^{-2}$ are more stable
than the two-quanta dislocation with $E_{2}\sim 2\pi B(2d/4\pi )^{4}\xi ^{-2}$.

\bigskip

Igor A. Luk'yanchuk

\smallskip

Institut f\"{u}r Theoretische Physik, RWTH-Aachen,

Templergraben 55, 52056 Aachen, Germany;

and

L.D.Landau Institute for Theoretical Physics, Moscow,

Russia

\bigskip




\begin{references}
\bibitem{Lub}  R. D. Kamien and T. C. Lubensky, Phys. Rev. Lett., {\bf 82},
2892 (1999)

\bibitem{Kleman}  M. Kl\'{e}man, {\it Point, Lines and Walls}:{\it \ in
Liquid Crystals, Magnetic Systems \ and Various Ordered Media }(Wiley, New
York, 1983)

\bibitem{TGBtheor}  S. R. Renn and T. C. Lubensky, Phys. Rev. {\bf A38},
2132 (1988)

\end{references}
\end{document}